\documentclass[11pt,a4paper,superscriptaddress]{revtex4-1}
\usepackage{graphicx,color}
\usepackage[top=30mm,bottom=30mm,left=30mm,right=30mm]{geometry}
\usepackage{placeins}
\usepackage{caption}
\usepackage{amssymb}
\usepackage[normalem]{ulem}
\usepackage{subcaption}
\usepackage{lineno}
\usepackage{indentfirst}
\usepackage{multirow}

\newcommand\onepot{\ensuremath{7.8\times 10^{21}\;\mbox{protons-on-target}} }
\newcommand\twopot{\ensuremath{20\times 10^{21}\;\mbox{protons-on-target}} }
\newcommand\nupot{\ensuremath{6.6\times 10^{20}\;\mbox{protons-on-target}} }

\newcommand\onepott{\ensuremath{7.8\times 10^{21}~\mbox{POT}} }
\newcommand\twopott{\ensuremath{20\times 10^{21}~\mbox{POT}} }

\graphicspath{{/figure}}
\begin{document}
\title{Sensitivity of the T2K accelerator-based neutrino
    experiment \\with an Extended run to $20\times10^{21}$ POT}

\newcommand{\INSTEE}{\affiliation{University of Bern, Albert Einstein Center for Fundamental Physics, Laboratory for High Energy Physics (LHEP), Bern, Switzerland}}
\newcommand{\INSTFE}{\affiliation{Boston University, Department of Physics, Boston, Massachusetts, U.S.A.}}
\newcommand{\INSTD}{\affiliation{University of British Columbia, Department of Physics and Astronomy, Vancouver, British Columbia, Canada}}
\newcommand{\INSTGA}{\affiliation{University of California, Irvine, Department of Physics and Astronomy, Irvine, California, U.S.A.}}
\newcommand{\INSTI}{\affiliation{IRFU, CEA Saclay, Gif-sur-Yvette, France}}
\newcommand{\INSTGB}{\affiliation{University of Colorado at Boulder, Department of Physics, Boulder, Colorado, U.S.A.}}
\newcommand{\INSTFG}{\affiliation{Colorado State University, Department of Physics, Fort Collins, Colorado, U.S.A.}}
\newcommand{\INSTFH}{\affiliation{Duke University, Department of Physics, Durham, North Carolina, U.S.A.}}
\newcommand{\INSTBA}{\affiliation{Ecole Polytechnique, IN2P3-CNRS, Laboratoire Leprince-Ringuet, Palaiseau, France }}
\newcommand{\INSTEF}{\affiliation{ETH Zurich, Institute for Particle Physics, Zurich, Switzerland}}
\newcommand{\INSTEG}{\affiliation{University of Geneva, Section de Physique, DPNC, Geneva, Switzerland}}
\newcommand{\INSTDG}{\affiliation{H. Niewodniczanski Institute of Nuclear Physics PAN, Cracow, Poland}}
\newcommand{\INSTCB}{\affiliation{High Energy Accelerator Research Organization (KEK), Tsukuba, Ibaraki, Japan}}
\newcommand{\INSTED}{\affiliation{Institut de Fisica d'Altes Energies (IFAE), The Barcelona Institute of Science and Technology, Campus UAB, Bellaterra (Barcelona) Spain}}
\newcommand{\INSTEC}{\affiliation{IFIC (CSIC \& University of Valencia), Valencia, Spain}}
\newcommand{\INSTEI}{\affiliation{Imperial College London, Department of Physics, London, United Kingdom}}
\newcommand{\INSTGF}{\affiliation{INFN Sezione di Bari and Universit\`a e Politecnico di Bari, Dipartimento Interuniversitario di Fisica, Bari, Italy}}
\newcommand{\INSTBE}{\affiliation{INFN Sezione di Napoli and Universit\`a di Napoli, Dipartimento di Fisica, Napoli, Italy}}
\newcommand{\INSTBF}{\affiliation{INFN Sezione di Padova and Universit\`a di Padova, Dipartimento di Fisica, Padova, Italy}}
\newcommand{\INSTBD}{\affiliation{INFN Sezione di Roma and Universit\`a di Roma ``La Sapienza'', Roma, Italy}}
\newcommand{\INSTEB}{\affiliation{Institute for Nuclear Research of the Russian Academy of Sciences, Moscow, Russia}}
\newcommand{\INSTHA}{\affiliation{Kavli Institute for the Physics and Mathematics of the Universe (WPI), The University of Tokyo Institutes for Advanced Study, University of Tokyo, Kashiwa, Chiba, Japan}}
\newcommand{\INSTCC}{\affiliation{Kobe University, Kobe, Japan}}
\newcommand{\INSTCD}{\affiliation{Kyoto University, Department of Physics, Kyoto, Japan}}
\newcommand{\INSTEJ}{\affiliation{Lancaster University, Physics Department, Lancaster, United Kingdom}}
\newcommand{\INSTFC}{\affiliation{University of Liverpool, Department of Physics, Liverpool, United Kingdom}}
\newcommand{\INSTFI}{\affiliation{Louisiana State University, Department of Physics and Astronomy, Baton Rouge, Louisiana, U.S.A.}}
\newcommand{\INSTJ}{\affiliation{Universit\'e de Lyon, Universit\'e Claude Bernard Lyon 1, IPN Lyon (IN2P3), Villeurbanne, France}}
\newcommand{\INSTHB}{\affiliation{Michigan State University, Department of Physics and Astronomy,  East Lansing, Michigan, U.S.A.}}
\newcommand{\INSTCE}{\affiliation{Miyagi University of Education, Department of Physics, Sendai, Japan}}
\newcommand{\INSTDF}{\affiliation{National Centre for Nuclear Research, Warsaw, Poland}}
\newcommand{\INSTFJ}{\affiliation{State University of New York at Stony Brook, Department of Physics and Astronomy, Stony Brook, New York, U.S.A.}}
\newcommand{\INSTGJ}{\affiliation{Okayama University, Department of Physics, Okayama, Japan}}
\newcommand{\INSTCF}{\affiliation{Osaka City University, Department of Physics, Osaka, Japan}}
\newcommand{\INSTGG}{\affiliation{Oxford University, Department of Physics, Oxford, United Kingdom}}
\newcommand{\INSTBB}{\affiliation{UPMC, Universit\'e Paris Diderot, CNRS/IN2P3, Laboratoire de Physique Nucl\'eaire et de Hautes Energies (LPNHE), Paris, France}}
\newcommand{\INSTGC}{\affiliation{University of Pittsburgh, Department of Physics and Astronomy, Pittsburgh, Pennsylvania, U.S.A.}}
\newcommand{\INSTFA}{\affiliation{Queen Mary University of London, School of Physics and Astronomy, London, United Kingdom}}
\newcommand{\INSTE}{\affiliation{University of Regina, Department of Physics, Regina, Saskatchewan, Canada}}
\newcommand{\INSTGD}{\affiliation{University of Rochester, Department of Physics and Astronomy, Rochester, New York, U.S.A.}}
\newcommand{\INSTHC}{\affiliation{Royal Holloway University of London, Department of Physics, Egham, Surrey, United Kingdom}}
\newcommand{\INSTBC}{\affiliation{RWTH Aachen University, III. Physikalisches Institut, Aachen, Germany}}
\newcommand{\INSTFB}{\affiliation{University of Sheffield, Department of Physics and Astronomy, Sheffield, United Kingdom}}
\newcommand{\INSTDI}{\affiliation{University of Silesia, Institute of Physics, Katowice, Poland}}
\newcommand{\INSTEH}{\affiliation{STFC, Rutherford Appleton Laboratory, Harwell Oxford,  and  Daresbury Laboratory, Warrington, United Kingdom}}
\newcommand{\INSTCH}{\affiliation{University of Tokyo, Department of Physics, Tokyo, Japan}}
\newcommand{\INSTBJ}{\affiliation{University of Tokyo, Institute for Cosmic Ray Research, Kamioka Observatory, Kamioka, Japan}}
\newcommand{\INSTCG}{\affiliation{University of Tokyo, Institute for Cosmic Ray Research, Research Center for Cosmic Neutrinos, Kashiwa, Japan}}
\newcommand{\INSTGI}{\affiliation{Tokyo Metropolitan University, Department of Physics, Tokyo, Japan}}
\newcommand{\INSTF}{\affiliation{University of Toronto, Department of Physics, Toronto, Ontario, Canada}}
\newcommand{\INSTB}{\affiliation{TRIUMF, Vancouver, British Columbia, Canada}}
\newcommand{\INSTG}{\affiliation{University of Victoria, Department of Physics and Astronomy, Victoria, British Columbia, Canada}}
\newcommand{\INSTDJ}{\affiliation{University of Warsaw, Faculty of Physics, Warsaw, Poland}}
\newcommand{\INSTDH}{\affiliation{Warsaw University of Technology, Institute of Radioelectronics, Warsaw, Poland}}
\newcommand{\INSTFD}{\affiliation{University of Warwick, Department of Physics, Coventry, United Kingdom}}
\newcommand{\INSTGE}{\affiliation{University of Washington, Department of Physics, Seattle, Washington, U.S.A.}}
\newcommand{\INSTGH}{\affiliation{University of Winnipeg, Department of Physics, Winnipeg, Manitoba, Canada}}
\newcommand{\INSTEA}{\affiliation{Wroclaw University, Faculty of Physics and Astronomy, Wroclaw, Poland}}
\newcommand{\INSTH}{\affiliation{York University, Department of Physics and Astronomy, Toronto, Ontario, Canada}}

\INSTEE
\INSTFE
\INSTD
\INSTGA
\INSTI
\INSTGB
\INSTFG
\INSTFH
\INSTBA
\INSTEF
\INSTEG
\INSTDG
\INSTCB
\INSTED
\INSTEC
\INSTEI
\INSTGF
\INSTBE
\INSTBF
\INSTBD
\INSTEB
\INSTHA
\INSTCC
\INSTCD
\INSTEJ
\INSTFC
\INSTFI
\INSTJ
\INSTHB
\INSTCE
\INSTDF
\INSTFJ
\INSTGJ
\INSTCF
\INSTGG
\INSTBB
\INSTGC
\INSTFA
\INSTE
\INSTGD
\INSTHC
\INSTBC
\INSTFB
\INSTDI
\INSTEH
\INSTCH
\INSTBJ
\INSTCG
\INSTGI
\INSTF
\INSTB
\INSTG
\INSTDJ
\INSTDH
\INSTFD
\INSTGE
\INSTGH
\INSTEA
\INSTH

\author{K.\,Abe}\INSTBJ
\author{C.\,Andreopoulos}\INSTEH\INSTFC
\author{M.\,Antonova}\INSTEB
\author{S.\,Aoki}\INSTCC
\author{A.\,Ariga}\INSTEE
\author{D.\,Autiero}\INSTJ
\author{S.\,Ban}\INSTCD
\author{M.\,Barbi}\INSTE
\author{G.J.\,Barker}\INSTFD
\author{G.\,Barr}\INSTGG
% ...

\collaboration{The T2K Collaboration}\noaffiliation

\date{\today}

\begin{abstract}
\vspace{5cm}
  Recent measurements at the T2K experiment
  indicate that CP violation in neutrino mixing may be observed
  in the future by long-baseline neutrino oscillation experiments. 
We explore the physics program
of an extension to the currently approved T2K running of
$7.8\times 10^{21}$ protons-on-target to $20\times 10^{21}$ protons-on-target,
aiming at initial observation of CP violation with 3$\,\sigma$
  or higher significance for the case of maximum CP violation.
With accelerator and beam line 
upgrades, as well as analysis improvements, this program would occur before the next generation of long-baseline neutrino oscillation experiments that are expected 
to start operation in 2026.
\end{abstract}

\maketitle

\section{Introduction}
The discovery of $\nu_\mu\to \nu_e$ oscillations
by the T2K accelerator-based long-baseline experiment\cite{Abe:2011ks, Abe:2013hdq} 
has opened the possibility of observing CP-violation (CPV) in the lepton sector,
which would be a crucial hint towards understanding the
matter-antimatter asymmetry of the universe\cite{Fukugita:1986hr}. In neutrino oscillations, CPV 
can arise from $\delta_{CP}$,  an  irreducible CP-odd phase
in the lepton mixing matrix,
which can be measured at T2K by comparing the $\nu_\mu\to\nu_e$ 
and $\bar{\nu}_\mu\to\bar{\nu}_e$ oscillation probabilities or by comparing 
these oscillations with $\bar{\nu}_e$ disappearance measured by reactors\cite{An:2016bvr,Ahn:2012nd,Abe:2014bwa}.
While the current significance is marginal, 
T2K measurements with $\nupot$ (POT) hint at maximum  CP violation with 
$\delta_{CP}\sim-\frac{\pi}{2}$ and normal mass hierarchy\cite{Abe:2015awa}.
Recent results from the NOvA experiment\cite{Adamson:2016tbq}, another accelerator-based long-baseline experiment, are 
consistent with this picture, though the statistical uncertainties are still large.
In this maximal case, T2K could observe CPV with 90\% C.L. sensitivity with the
$\onepott$ currently approved by J-PARC and expected by around 2020\cite{Abe:2014tzr}.
Future proposed projects such as Hyper-Kamiokande\cite{Abe:2015zbg} 
and DUNE\cite{dune} aim to achieve $>3\;\sigma$ sensitivity to CPV across a 
wide range of  $\delta_{CP}$ on the time scale of 2026 and beyond.

By the time T2K finishes its currently approved running,
the J-PARC Main Ring (MR) beam 
power is expected to exceed $750$~kW. If data-taking is extended until 2026, 
when Hyper-Kamiokande and DUNE are expected to start, sensitivity to CPV would 
significantly improve with the additional statistics. This would also have 
the benefit of establishing higher beam power for the next generation of 
measurements at Hyper-Kamiokande from the start.

The T2K collaboration has initiated the study of ``T2K-II'', a second phase of 
the experiment in  which more than $3\,\sigma$ sensitivity to CPV can be achieved
if  $\delta_{CP}\sim-\frac{\pi}{2}$ and the mass hierarchy is normal
in a five or six year period
after the currently approved running.
This would require not only a beam time extension, but additional improvements 
explored in this document, including further improvements to the MR beam power,
 neutrino beam line upgrades, and analysis developments to improve statistical 
and systematic uncertainties. We discuss the physics potential resulting 
from these combined developments.

\section{Data accumulation Plan and Improvement of effective Statistics}
\paragraph{Projected MR beam power and POT accumulation}
The MR beam power has steadily increased since the start of the operation.
In May 2016, 420~kW beam with 2.2$\times10^{14}$ protons-per-pulse (ppp)
every 2.48 seconds was successfully provided to the neutrino beamline.
Discussions with the J-PARC Accelerator Group have resulted in a plan to achieve the
design intensity of 750~kW by reducing the repetition cycle to
1.3 seconds.
This requires an 
upgrade to the power supplies for the MR main magnets, RF cavities, and some
injection and extraction devices by January 2019. 
Studies to increase the ppp are also in progress, with
a $2.73\times 10^{14}$ ppp-equivalent beam
with acceptable beam loss already demonstrated in a test operation
with two bunches.

\begin{figure} \centering
\includegraphics[width=0.65\textwidth]{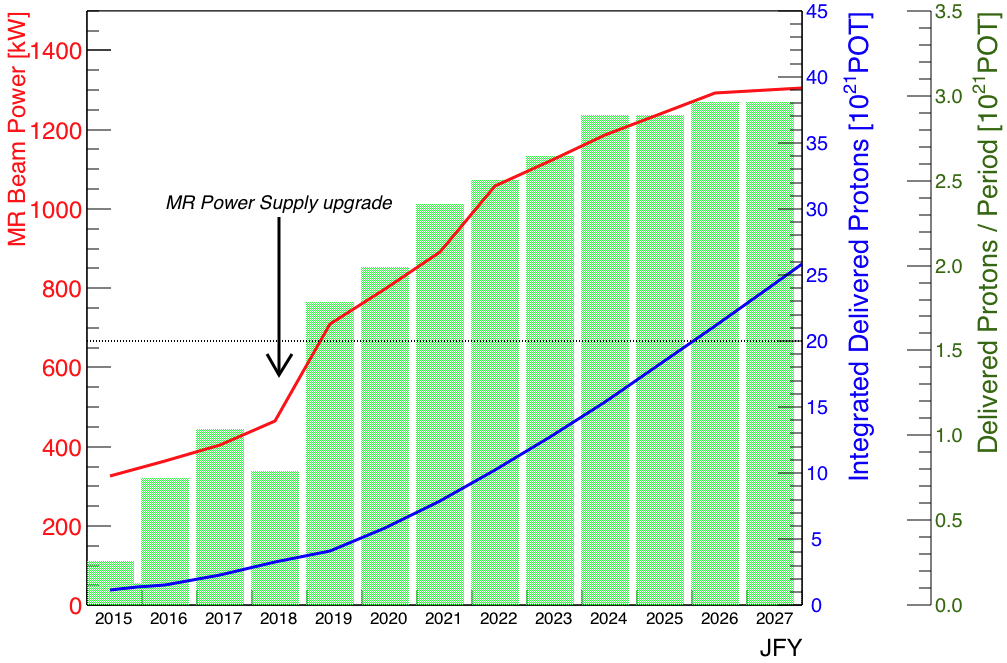}
\caption[POT]{Anticipated MR beam power and POT accumulation vs. calendar year.
\label{fig:POT}}
\end{figure}

Based on these developments,
MR beam power prospects were updated and presented 
in the accelerator report at the last PAC in July 2015\cite{pac16}
and anticipated beam power of 1.3~MW with 3.2$\times$10$^{14}$ ppp
and a repetition cycle of 1.16 seconds were presented\cite{jparcnuws,nnn15}.
A possible data accumulation scenario is shown in Fig.~\ref{fig:POT},
where 5 months of neutrino beam operation each year and realistic
running time efficiency are assumed.
We expect to accumulate $\twopott$ by JFY2026
with 5 months of operation each year and
by JFY2025 with 6 months of operation each year.

\paragraph{Beamline upgrade}
The beam intensity in the current neutrino beam facility 
is limited to $3.3\times 10^{14}$ ppp by the thermal shock induced by the beam on the target 
and beam window.
The MR power upgrade plan allows 1.3~MW beam operation without increasing the ppp. However, the beamline cooling capacity for components like the target
and helium vessel is sufficient for up to 750~kW;
these would need to be upgraded to accept 1.3 MW beam operation.

The T2K horns were originally designed to be operated at 320~kA current,
but so far have been operated at 250~kA because of a problem with the power supplies.
The upgrades required for 320~kA operation
will be implemented in stages and will be completed
by 2019. 
Horn operation at 320~kA gives a 10\% higher neutrino flux 
and also reduces contamination of the wrong-sign component of neutrinos
({\em i.e.}, anti-neutrinos in the neutrino beam mode
or neutrinos in the anti-neutrino beam mode) by 5-10\%.

\paragraph{Improved Super-K Sample Selection}
The current efficiency to select oscillated $\nu_e$ CC events in the 22.5 kt fiducial volume
at Super-K is 66\%.
The inefficiency results from targeting events with a single Cherenkov ring from the outgoing lepton without additional rings or decay electrons arising from pions that may be produced in the interaction.
Recent developments in multi-ring event reconstruction will enable us
to identify and reconstruct  $\nu_e$
CC $\pi^{\pm/0}$ interactions,
leading to higher effective efficiency for the $\nu_e$ CC selection.
Reoptimization of other selection criteria are also being investigated.

Improvements to the single-ring $\mu$-like
selection used to identify $\nu_\mu$ CC events will enhance T2K's sensitivity to $\theta_{23}$ and $\Delta m_{32}^2$  and subsequently CP violation through the improved constraint on these parameters.
We expect to reduce the NC$\pi^+$ contamination in this sample by more than 50\% in the region where the oscillation effect is maximal.
As with the $\nu_e$, a dedicated multi-ring $\nu_{\mu}$ CC$\pi^{+}$ reconstruction is under development, potentially allowing up to 40\% more $\nu_\mu$ CC events to be used in the oscillation analyses.

Finally, the fiducial volume definition for both selections will be improved to accept well-reconstructed events near the edge of the detector that are currently rejected. This is expected to add 10-15\% more events while maintaining sufficient control of external backgrounds entering the tank.

Taken together, these improvements can potentially increase the $\nu_e$ and $\nu_\mu$ CC event samples identified at Super-K by up to 40\%.

\paragraph{Short Summary}
We expect to accumulate an integrated \twopott
when T2K running is extended by five to six years.
Effective statistics per POT for CP violation studies will be
improved by up to 50\% by analysis improvements and beamline upgrades. 

The number of events expected at the Super-K far detector
  for an exposure of \(20\times10^{21}\) POT with a 50\% statistical
  improvement is given in Table \ref{tab:detevts} assuming either true
  \(\delta_{CP} = 0\) or \(-\pi / 2\). 
\begin{table}
\begin{center}
\caption[Number of Expected Events]{Number of events expected to be observed
        at the far detector for
\(10\times10^{21}\)~POT \(\nu\)- + \(10\times10^{21}\)~POT \(\bar{\nu}\)-mode
with a 50\% statistical improvement.
Assumed relevant oscillation parameters are:
\(\sin^22\theta_{13}=0.085\), \(\sin^2\theta_{23}=0.5\), 
\(\Delta m^2_{32}=2.5\times10^{-3}\) eV\(^2\), and normal mass hierarchy (MH).
\label{tab:detevts}}
\begin{tabular}{c | c | c | c | c | c | c | c   } \hline
& & & Signal & Signal & Beam CC & Beam CC & \\
& True \(\delta_{CP}\) & Total & \(\nu_{\mu} \rightarrow \nu_e\) & \(\bar{\nu}_{\mu} \rightarrow
\bar{\nu}_e\) & \(\nu_e + \bar{\nu}_e \) & \(\nu_{\mu} + \bar{\nu}_{\mu} \) & NC\\ \hline\hline
\(\nu\)-mode & 0  & 467.6 & 356.3 &  4.0 & 73.3 & 1.8 & 32.3 \\ \cline{2-8}
$\nu_e$ sample & \(-\pi/2\) & 558.7 & 448.6 &  2.8 & 73.3 & 1.8 & 32.3 \\ \hline \hline
\(\bar{\nu}\)-mode & 0          & 133.9 & 16.7 &  73.6 & 29.2 & 0.4 & 14.1 \\ \cline{2-8}
$\bar{\nu}_e$ sample & \(-\pi/2\) & 115.8 & 19.8 &  52.3 & 29.2 & 0.4 & 14.1 \\ \hline 
\end{tabular} 
\vskip 0.4cm
\begin{tabular}{  c | c | c | c | c | c | c   } \hline
& & Beam CC & Beam CC & Beam CC & \(\nu_{\mu} \rightarrow \nu_e +\) & \\
& Total & \(\nu_\mu\) & \(\bar{\nu}_\mu\) & \(\nu_e + \bar{\nu}_e \) & \(\bar{\nu}_{\mu} \rightarrow \bar{\nu}_e\) & NC\\ \hline\hline
\(\nu\)-mode $\nu_\mu$ sample & 2735.0 & 2393.0 &  158.2 & 1.6 & 7.2 & 175.0 \\ \hline \hline
\(\bar{\nu}\)-mode $\bar{\nu}_\mu$ sample & 1283.5 & 507.8 &  707.9 & 0.6 & 1.0 & 66.2 \\ \hline
\end{tabular}
\end{center}
\end{table}


\section{Improvement of Systematics}
\label{sec:anaimp}

Systematic errors are categorized based on their source into  
neutrino flux, neutrino interaction model, and detector model uncertainties.
The uncertainties in the neutrino flux and 
interaction model are first constrained by external measurements
and then further constrained by a fit to data from the ND280 near detector.

The uncertainty on the total predicted number of events
in the Super-K samples encapsulates the first order impact of
systematic errors on the oscillation parameter measurements
and the current sizes are summarized
in Table~\ref{tab:syst_sum}.
The CP phase $\delta_{CP}$ is measured
through the difference in the oscillation probabilities for $\nu_\mu\to\nu_{e}$
and $\bar{\nu}_\mu\to\bar{\nu}_{e}$.
Hence, we also show the uncertainty on the ratio of expected $\nu_e/\bar{\nu}_e$ candidates
at Super-K with neutrino ($\nu$) and antineutrino ($\bar{\nu}$) beam mode.

The uncertainty from oscillation parameters not measured by T2K-II 
is negligible for $\nu_\mu/\bar{\nu}_\mu$ events
at SK in the $\nu_\mu/\bar{\nu}_\mu$ disappearance measurements.
The 4\% uncertainties on the $\nu_e/\bar{\nu}_e$ samples arise mainly
from the precision of the $\theta_{13}$ measurement by reactor experiments($\sin^{2}(2\theta_{13})=0.085\pm0.005$)\cite{pdg}.
However, this uncertainty is  correlated between $\nu$ and $\bar{\nu}$ beam 
mode samples and its impact on the observation of a CP asymmetry in T2K data
is small.

As will be described in Sec.~\ref{sec:physics},
the current systematic errors, if they are not improved,
will significantly reduce the sensitivity
to CP violation with the T2K-II statistics.
Any improvement on the systematics would enhance physics potential.
Here, we describe projected improvements.

\begin{table}[t]
 \caption{Errors on the number of predicted events
 in the Super-K samples from individual 
systematic error sources in neutrino ($\nu$ mode) and
 antineutrino beam mode ($\bar{\nu}$ mode).
Also shown is the error on the ratio 1R$e$ events in $\nu$ mode/$\bar{\nu}$ mode. The uncertainties represent work-in-progress for T2K neutrino oscillation results in 2016.}
 \label{tab:syst_sum}
 \begin{center}
   \begin{tabular}{l|c|c|c|c|c}
     \hline \hline
                  & \multicolumn{5}{|c}{$\delta_{N_{SK}}/N_{SK}$ (\%)}  \\ \hline 
                  &  \multicolumn{2}{|c}{ 1-Ring $\mu$} & \multicolumn{3}{|c}{ 1-Ring $e$} \\ \hline
     Error Type   & $\nu$ mode & $\bar{\nu}$ mode & $\nu$ mode & $\bar{\nu}$ mode & $\nu$/$\bar{\nu}$ \\ \hline \hline
     SK Detector  &  4.6    & 3.9    & 2.8    &  4.0   & 1.9     \\ \hline
     SK Final State \& Secondary Interactions  & 1.8   & 2.4  &  2.6  & 2.7  & 3.7     \\ \hline
     ND280 Constrained Flux \& Cross-section   & 2.6  & 3.0 & 3.0 & 3.5 & 2.4     \\ \hline 
     $\sigma_{\nu_{e}}/\sigma_{\nu_{\mu}}$, $\sigma_{\bar{\nu}_{e}}/\sigma_{\bar{\nu}_{\mu}}$ &  0.0  & 0.0   & 2.6 & 1.5 & 3.1     \\ \hline
     NC 1$\gamma$ Cross-section &  0.0  & 0.0 & 1.4 & 2.7 & 1.2     \\ \hline
     NC Other Cross-section     &  0.7  & 0.7 & 0.2 & 0.3 & 0.1     \\ \hline \hline
     Total Systematic Error     &  5.6  & 5.5 & 5.7 & 6.8 & 5.9    \\ \hline \hline
     External Constraint on $\theta_{12}$, $\theta_{13}$, $\Delta m^{2}_{21}$  & 0.0 & 0.0 & 4.2 & 4.0 & 0.1     \\
     \hline \hline
   \end{tabular}
 \end{center}
\end{table}


\paragraph{Neutrino Flux}

The neutrino flux prediction\cite{Abe:2012av} uncertainty is currently dominated by uncertainties
on the hadron interaction modelling in the target and surrounding materials in the neutrino beamline
and by the proton beam orbit measurement.
These errors  can be represented as an absolute flux uncertainty relevant for neutrino cross section measurements,
and an extrapolation uncertainty which impacts oscillation measurements.
At the peak energy ($\sim 600$ MeV), these are currently $\sim 9\%$
and $\sim0.3\%$
, respectively. Further improvement is expected with the incorporation of the T2K replica target data from NA61/SHINE, improvements in the beam direction measurement, and improved usage of the near detector measurements, to achieve $\sim 6\%$ uncertainty on the absolute flux.

\paragraph{Near Detector measurement}

Currently, detector-related systematic uncertainties of $\sim2\%$ have been
achieved in $\nu_\mu/\bar{\nu}_\mu$ charged-current samples selected in ND280.
Some uncertainties, such as those related to
reconstruction efficiencies and backgrounds, may be reduced by further effort and development. By far the largest uncertainty, however, arises from pion secondary interaction uncertainties, which may be reduced by external measurements or by studying pion interactions within ND280 itself. With additional data, we expect to reduce this uncertainty and achieve $\sim1\%$ overall systematic error in the ND280 samples.

\paragraph{Neutrino Interaction}
T2K has engaged in continuous development and improvement of neutrino-nucleus interaction modelling\cite{Hayato:2009zz,Wilkinson:2016wmz}, including effects arising from nucleon correlations\cite{Martini:2009,Nieves:2011yp} and final state interaction of hadrons within the target nucleus. These models are further constrained by the near detector, but the constraints 
are limited by differences in the neutrino energy spectrum and acceptances between the near detector and Super-K.

We will continue to engage in model developments and comparisons with ND280 and externally published measurements.
Combined with the recent incorporation of neutrino interactions on the water targets and future improvements to the phase space coverage of the ND280 measurements, systematic errors, and flux prediction uncertainties, we expect to reduce the flux and cross section systematics. The large sample of $\nu_e/\bar{\nu}_e$ events in ND280 with the additional running will also allow us to improve the uncertainties arising from uncertainties in the ratios $\sigma_{\nu_e}/\sigma_{\nu_\mu}$ and $\sigma_{\bar{\nu}_e}/\sigma_{\bar{\nu}_\mu}$\cite{Day:2012gb}.
In addition, a task force was formed by the collaboration in 2015 to investigate the prospect and need of ND280 upgrade.

\paragraph{Super-K Systematics Improvement}
The current Super-K detector systematic errors are determined mainly by a fit
to the Super-K atmospheric neutrino data
and constraints on the energy scale uncertainty from cosmic muon control samples. The atmospheric neutrino fit will be updated to include the cross section
modelling from the T2K data.
Longer-term improvements would utilize calibration, entering muon, and decay electron data to constrain fundamental detector parameters, rather than fitting neutrino data, which is susceptible to atmospheric flux and neutrino cross section uncertainties. The expected improvement to the Super-K detector uncertainties is under study. The secondary interaction and final state interaction systematic errors uncertainties will also benefit from the ND280 and external pion interaction measurements and neutrino interaction model development.

\paragraph{Short Summary}
The current systematic error on the far detector prediction is
5.5 to 6.8\%. Considering the present situation and projected
improvements, we consider that 4\% systematic error is a reachable
and reasonable target for T2K-II. In what follows, this improvement in systematic error is modelled
by scaling the covariance matrix that reflects the current systematic error to obtain an uncertainty in the far detector prediction
that is 2/3 its current size.
Whether a near detector upgrade is needed to achieve this goal
will be investigated in one year time scale.
\section{Expected Physics Outcomes}
\label{sec:physics}
\paragraph{CP violation and precise determination of $\Delta m^2_{32}$ and
$\sin^2\theta_{23}$}\ \\

We assume that the full T2K-II exposure is 
\(20\times10^{21}\) POT taken equally in \(\nu\)-mode and \(\bar{\nu}\)-mode.
Further optimization of the running ratio between \(\nu\)-mode
and \(\bar{\nu}\)-mode will be pursued in the future.
Sensitivities were initially calculated with the current T2K (2016 oscillation analysis) event rates and systematics, and the effect of the enhancements from beam line and analysis improvements was implemented by a simple scaling.
Assumed relevant oscillation parameters are:
\(\sin^22\theta_{13}=0.085\), \(\sin^2\theta_{23}=0.5\), 
\(\Delta m^2_{32}=2.5\times10^{-3}\) eV\(^2\), and normal mass hierarchy (MH).
Cases for the current 90\% C.L. edges of \(\sin^2\theta_{23}\)
{\it i.e.} 0.43 and 0.6 are also studied.

\begin{figure} \centering 
\begin{subfigure}[t]{7.2cm}
\includegraphics[width=7.2cm]{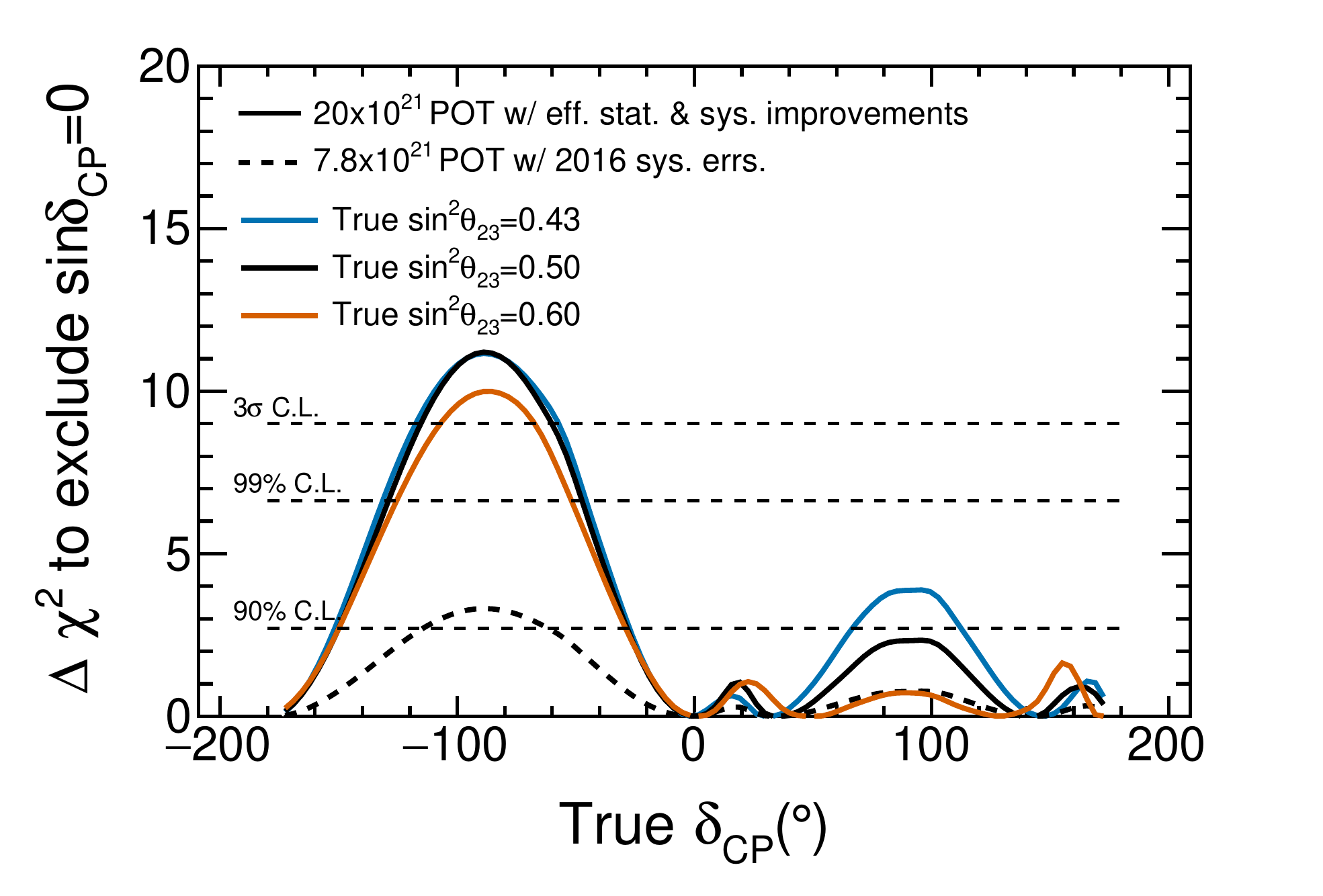}\caption{Assuming the MH is unknown.} \label{fig:CPVvsdCP_unknownMH}
\end{subfigure} \quad 
\begin{subfigure}[t]{7.2cm}
\includegraphics[width=7.2cm]{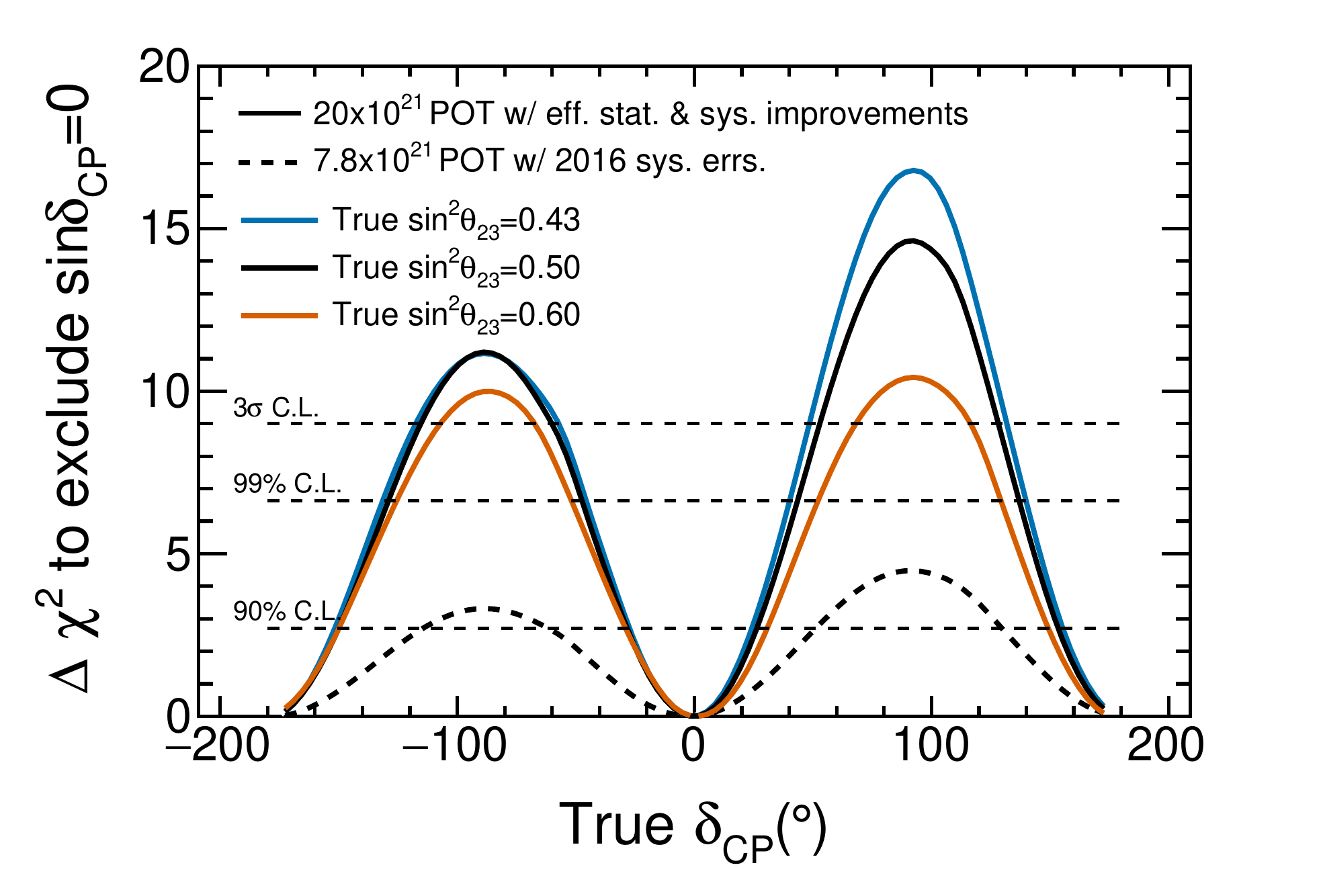}\caption{Assuming the MH is known -- measured by an outside experiment.} \label{fig:CPVvsdCP_knownMH}
\end{subfigure} \quad
\caption[CPV vs dCP]{Sensitivity to CP violation as a function of true
\(\delta_{CP}\) for the full T2K-II exposure of $20\times 10^{21}$ POT
with a 50\% improvement in the effective statistics, a reduction of the systematic uncertainties to 2/3 of their current size, and
assuming that the  true MH is the normal MH.
\label{fig:CPVvsdCP}} \end{figure}

The sensitivity to CP violation (\(\Delta\chi^2\) for resolving \(\sin\delta_{CP}\neq0\)) plotted 
as a function of true \(\delta_{CP}\) is given in Fig.\ \ref{fig:CPVvsdCP} 
for the full T2K-II exposure with a 50\% statistical improvement
and a reduction of the systematic uncertainties to 2/3 of its current magnitude.
When calculating sensitivities, the values of \(\sin^2\theta_{23}\), \(\Delta m^2_{32}\), and \(\delta_{CP}\) are
assumed to be constrained by the T2K-II data only, while \(\sin^22\theta_{13}\) is constrained by 
\(\sin^22\theta_{13}=0.085\pm0.005\)\cite{pdg}.

Several experiments (JUNO\cite{An:2015jdp}, NOvA\cite{Patterson:2012zs}, ORCA\cite{Katz:2014tta},
PINGU\cite{Aartsen:2014oha}) are expected or plan to determine the mass hierarchy before or during
the proposed period of T2K-II.
Hence both MH-unknown and -known cases are shown in Fig.~\ref{fig:CPVvsdCP}.
The fractional region for which \(\sin\delta_{CP}=0\) can be excluded
at the 99\% (3\(\sigma\)) C.L.\ is 49\% (36\%) of possible true values of \(\delta_{CP}\) assuming the improved systematic errors 
and that the MH has been determined by an outside experiment.
If systematic errors are eliminated completely, the 
fractional region where CPV can be resolved by 99\% (3\(\sigma\))
becomes 51\% (43\%). More details of coverage at different values of $\sin^2\theta_{23}$ can be found in Table~\ref{tab:coverage}.
\begin{table}[th]
  \caption{Table of $\delta_{CP}$ fractional coverages (\%) with three options of systematic treatment: no systematic error (statistical only), 2016 systematics and improved systematics. The coverages are calculated at three different values of $\sin^2\theta_{23}$ (0.43, 0.5, and 0.60) and it is assumed that the MH has been determined by an outside experiment.}
  \label{tab:coverage}
  \centering
  \begin{tabular}{c|c|c|c|c|c|c}
    \hline
    \multirow{2}{*}{} & \multicolumn{2}{c|}{$\sin^2\theta_{23}=0.43$   } &\multicolumn{2}{c|}{$\sin^2\theta_{23}=0.50$   } &  \multicolumn{2}{c}{$\sin^2\theta_{23}=0.60$  } \\
    \hline{~-----}
    & 99$\%$ C. L. & 3$\sigma$ & 99$\%$ C. L. & 3$\sigma$  & 99$\%$ C. L.  & 3$\sigma$  \\
    \hline
    Stat. Only & 57.5 & 47.9 & 53.3 & 43.1 & 49.1 & 36.7\\ 
    \hline
    2016 systematics & 45.6 & 28.3 & 41.6 & 20.5 & 34.7 & 5.2\\ 
    \hline 
    Improved systematics & 51.5 & 39.7 & 48.6 & 36.1 & 41.8 & 23.9\\ 
    \hline
  \end{tabular}
  
\end{table}

\begin{figure} \centering
\begin{subfigure}[t]{7.2cm}
\includegraphics[width=7.2cm]{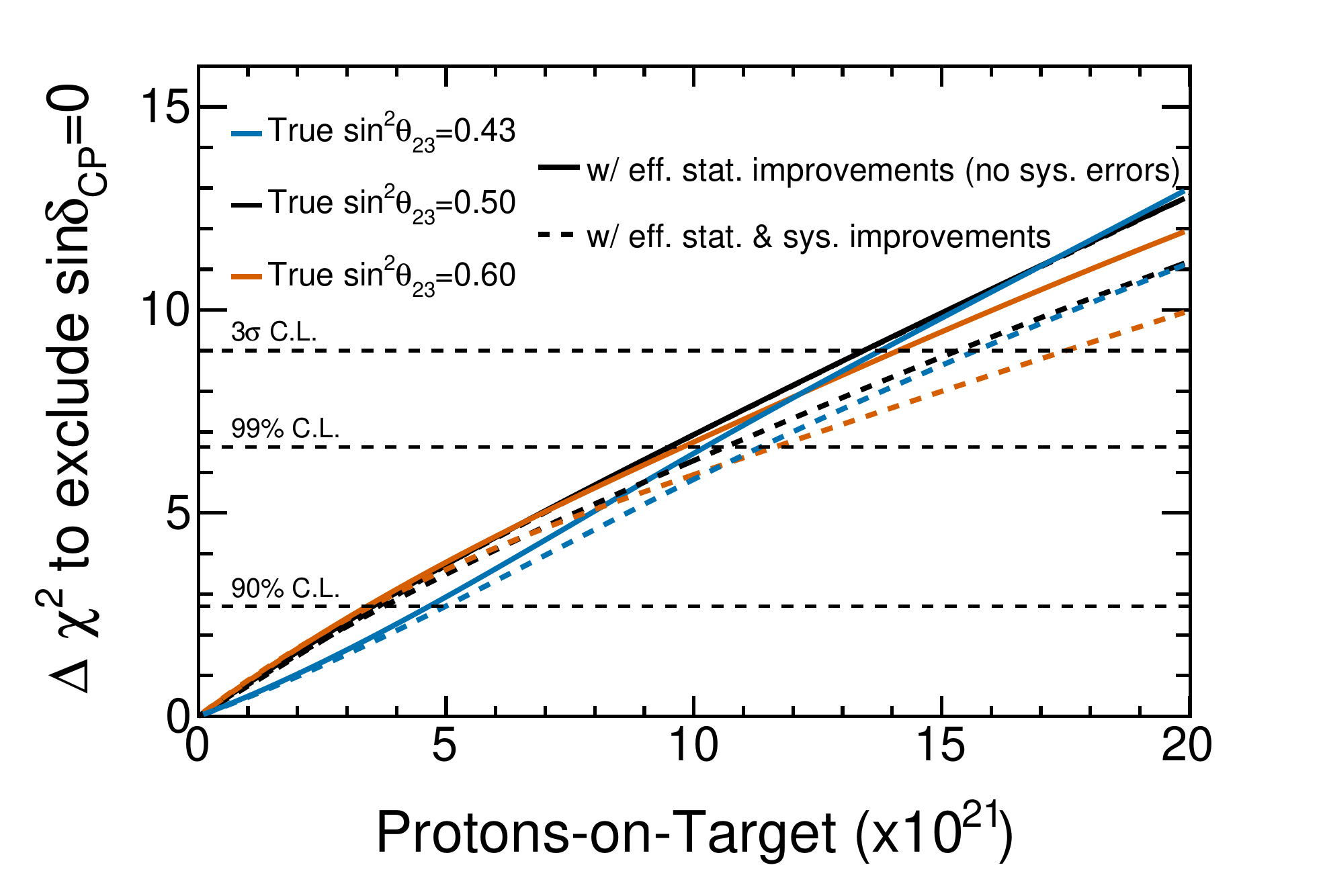}
\end{subfigure} \quad 
\begin{subfigure}[t]{7.2cm}
\includegraphics[width=7.2cm]{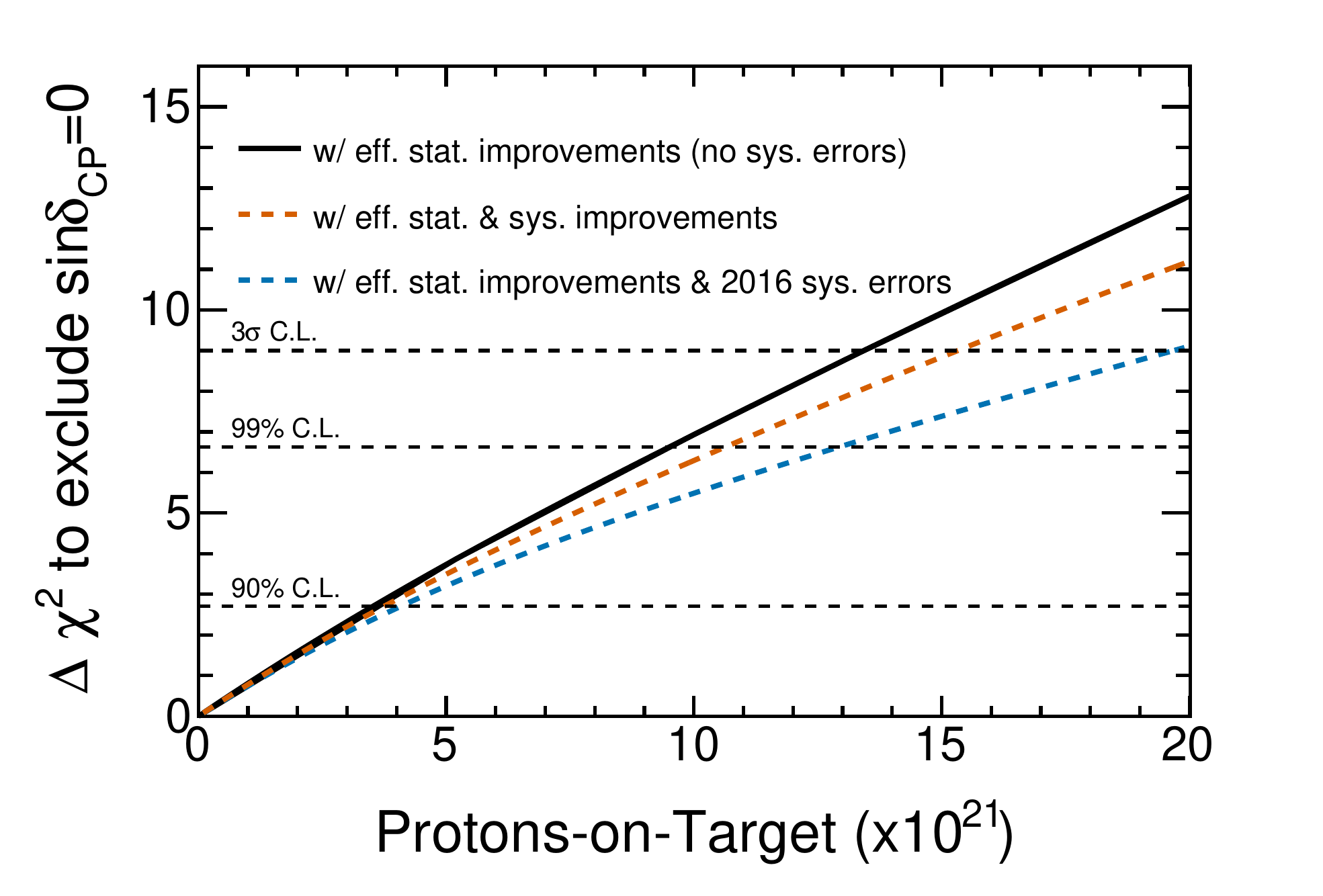}
\end{subfigure} \quad 
\caption[CPV vs POT]{Sensitivity to CP violation as a function of POT
with a 50\% improvement in the effective statistics,
assuming the true MH is the normal MH
and the true value of \(\delta_{CP}=-\pi/2\).
The plot on the left
compares different true values of \(\sin^2\theta_{23}\), while that on the
right compares different assumptions for the T2K-II systematic errors with $\sin^2\theta_{23}=0.50$.
\label{fig:CPVvsPOT}} \end{figure}

The expected evolution of the sensitivity to CP violation (\(\Delta\chi^2\)
for resolving \(\sin\delta_{CP}\neq0\)) as a function of POT assuming
that the T2K-II data is taken in roughly equal alternating periods of
\(\nu\)-mode and \(\bar{\nu}\)-mode (with true normal MH 
and \(\delta_{CP}=-\pi/2\)) is given in Fig.~\ref{fig:CPVvsPOT}.

\begin{figure} \centering
\begin{subfigure}[t]{7.2cm}
\includegraphics[width=7.2cm]{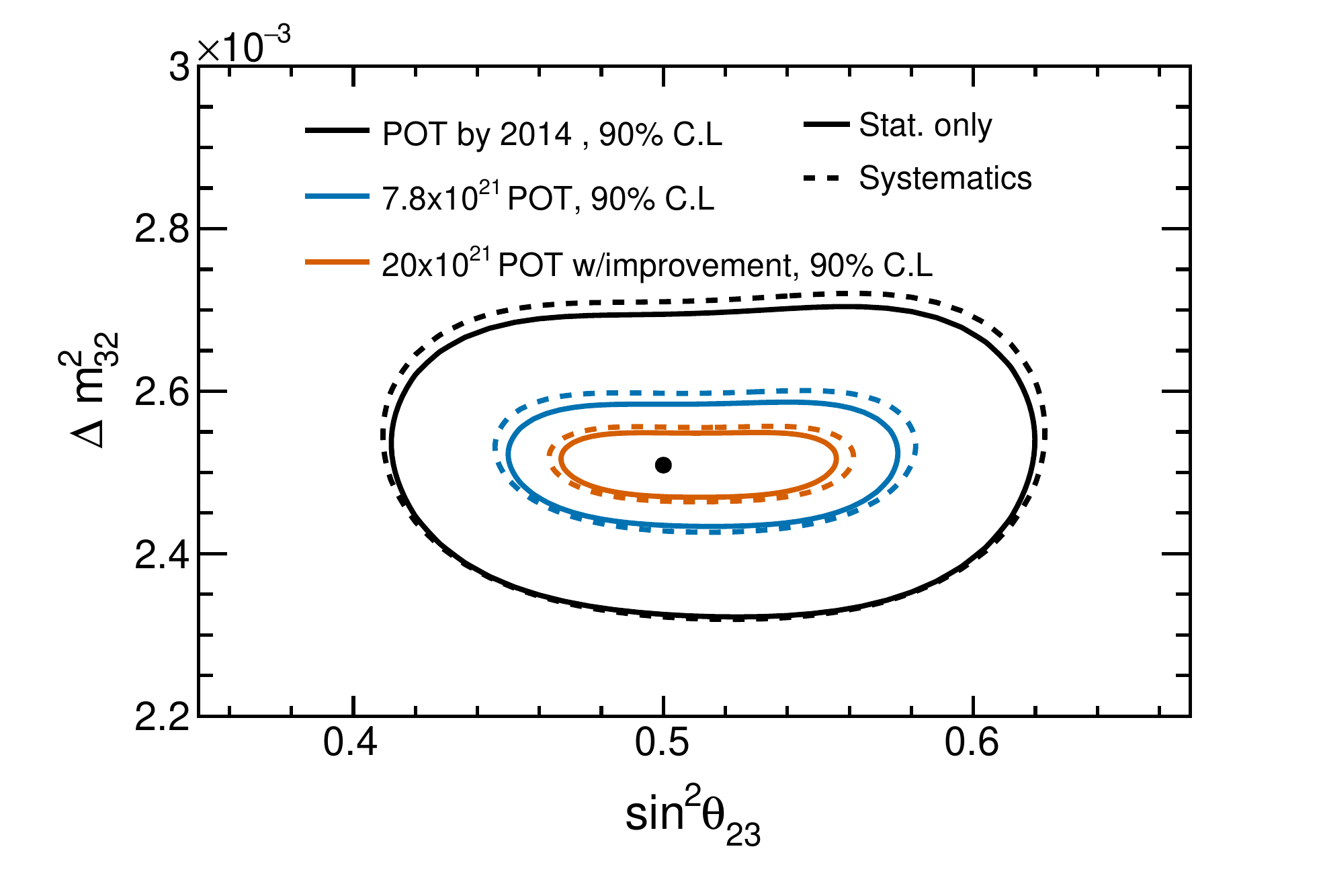}\caption{Assuming true \(\sin^2\theta_{23}=0.50\).}
\end{subfigure} \quad 
\begin{subfigure}[t]{7.2cm}
\includegraphics[width=7.2cm]{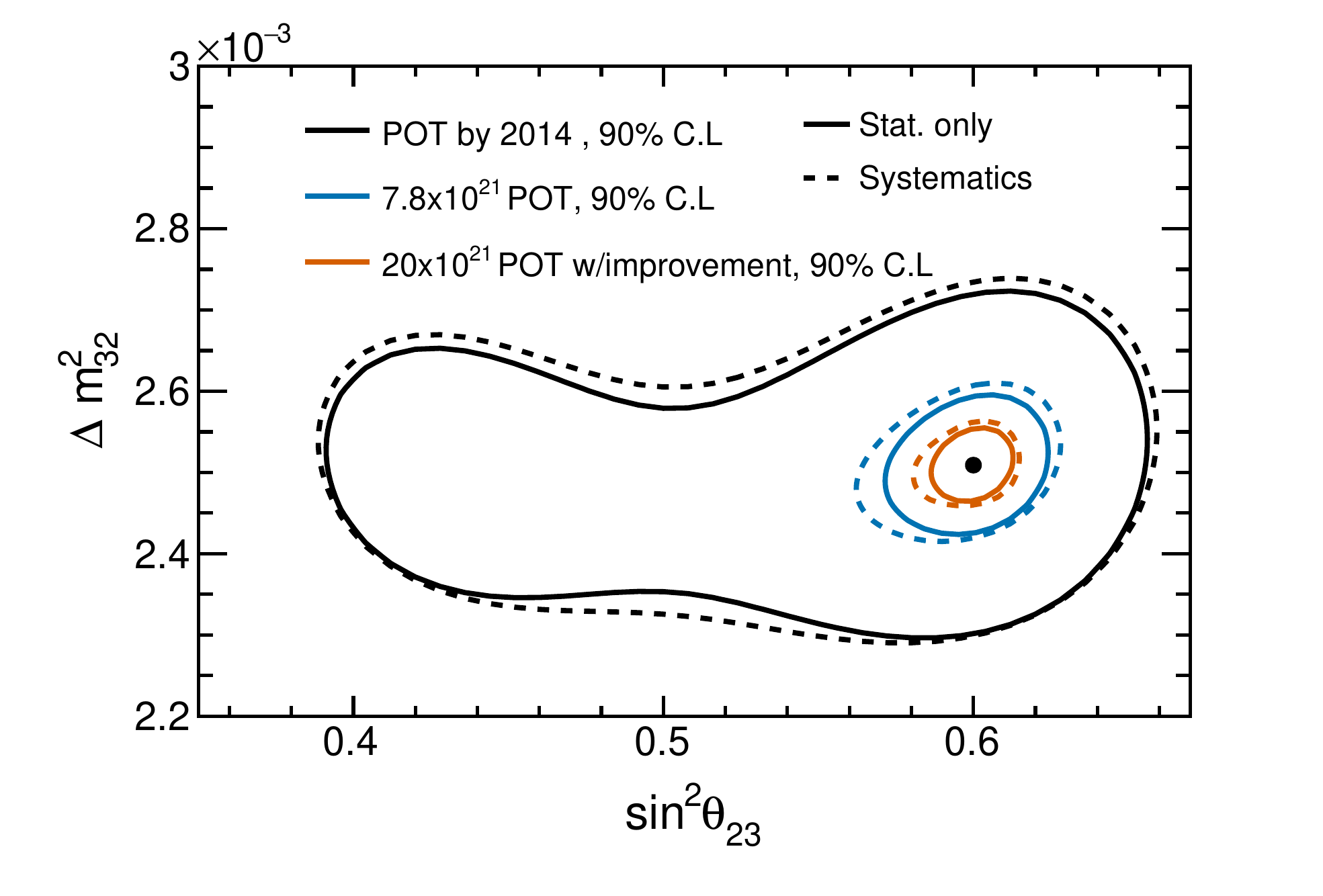}\caption{Assuming true \(\sin^2\theta_{23}=0.60\).}
\end{subfigure} \quad 
\caption[$\Delta m^2_{32}$ vs $\sin^2\theta_{23}$]{Expected 90\% C.L. sensitivity to $\Delta m^2_{32}$ and $\sin^2\theta_{23}$
with the 2016 systematic error.
The current POT corresponds to \(6.9\times10^{20}\) POT \(\nu\)-mode + \(4.0\times10^{20}\) POT \(\bar{\nu}\)-mode.
For the ultimate T2K-II exposure of $20\times 10^{21}$ POT, a 50\% increase in effective statistics is assumed.
\label{fig:dm2vss2th23}} \end{figure}

The expected 90\% C.L. contour for $\Delta m^2_{32}$ vs $\sin^2\theta_{23}$ for the full T2K-II exposure is shown in 
Fig.\ \ref{fig:dm2vss2th23}.
The expected 1\(\sigma\) precision on $\sin^2\theta_{23}$ is
\(\sim1.7^\circ(\sim0.7^\circ)\)
assuming \(\sin^2\theta_{23}=0.5\) (\(\sin^2\theta_{23}=0.43\), 0.6), and the expected precision 
on \(\Delta m^2_{32}\) is \(\sim\)1\%  assuming the true oscillation parameters
given above and true \(\delta_{CP}=-\pi/2\).

%

\paragraph{Neutrino Interaction Studies}\ \\
The additional run time of T2K-II will provide improved measurements
of neutrino and antineutrino scattering,
which probe nuclear structure through the axial vector current;
these data sets may be used to solve long-standing experimental disagreements
seen in previous measurements.
The reduced uncertainties of the neutrino/antineutrino flux,
increased statistical samples, and improvements to the acceptance of
the T2K detectors will enable more detailed kinematic measurements
to be made for interaction channels already measured by T2K,
including studies of nuclear effects relevant for quasi-elastic
and single pion resonance channels and measurements on water.
T2K also has near detectors placed in two different locations;
combined measurements of these detectors provide unique information
about the energy dependence of neutrino interactions.

With T2K-II, there are two opportunities
for neutrino interaction studies which are otherwise limited
by statistical uncertainty.
First are measurements of neutrino interactions in the argon gas of the TPCs,
where a very low threshold for tracking (below what can be achieved
with liquid argon detectors) can provide unique information
about proton multiplicity in neutrino-nucleus interactions. 
Approximately 10,600 $\nu$-Ar and 1,900 $\bar{\nu}$-Ar interactions
are expected.
Second, with expected datasets of 8,000 $\nu_e$ CC
and 2,000 $\bar{\nu}_e$ CC candidates, 
the differences between electron
and muon neutrino interactions can be studied;
these differences are an important source of systematic uncertainty
for CP violation measurements.

\paragraph{Non-standard Physics Studies}
\ \\
The high statistics at T2K-II would enable world-leading searches for various
physics beyond the standard model.
The combination of accelerator-based long-baseline measurements
with $\nu_\mu/\bar{\nu}_\mu$ beams and reactor measurements with
$\bar{\nu}_e$ flux may give redundant constraints
on ($\Delta m^2_{32}, \sin^2\theta_{23}, \delta_{CP}$).
Any inconsistency among these measurements
would indicate new physics such as unitarity violation
in the three-flavor mixing, sterile neutrinos,
non-standard interactions, or CPT violation.
With measurements at the near detectors,
one could search for, for example, sterile neutrinos
introduced to account for the LSND\cite{Aguilar:2001ty} or reactor anomalies\cite{Mention:2011rk},
non-standard interactions in neutrino production or interaction,
heavy sterile neutrino decay, and neutrino magnetic moments larger than
the standard model prediction.
Sidereal time dependence of the event rate either at the near detector
or Super-K can be used to search for Lorentz violation\cite{Kostelecky:2003cr}.

Since neutrino mass likely originates from
physics at very high energy scales ($\gtrsim 10^{14}$~GeV),
new physics at these energy scales could produce effects
of comparable size to neutrino oscillation.
Redundant and precise measurements
of neutrino oscillation are equally compelling and complementary
to precision searches at colliders for 
new physics at the TeV scale.

\section{Summary}
The prospect of the accelerator intensity and beamline upgrades togehter with analysis improvements
are discussed based on the running experience.
The extended running of the T2K experiment from \onepot to \twopot enables
exploration of CP violation in a wide range of $\delta_{CP}$ with 99\%C.L.,
to reach $3\,\sigma$ or higher sensitivity
for the case of maximum CP violation, to precisely determine
oscillation parameters, and to search for possible new physics.
This program would occur before the next generation of long-baseline
neutrino oscillation experiments that are expected to start operation in 2026.

\section*{Acknowledgment}
We thank the J-PARC staff for superb accelerator performance and the 
CERN NA61 Collaboration for providing valuable particle production data.
We acknowledge the support of MEXT, Japan; 
NSERC (Grant No. SAPPJ-2014-00031), NRC and CFI, Canada;
CEA and CNRS/IN2P3, France;
DFG, Germany; 
INFN, Italy;
National Science Centre (NCN), Poland;
RSF, RFBR and MES, Russia; 
MINECO and ERDF funds, Spain;
SNSF and SERI, Switzerland;
STFC, UK; and 
DOE, USA.
We also thank CERN for the UA1/NOMAD magnet, 
DESY for the HERA-B magnet mover system, 
NII for SINET4, 
the WestGrid and SciNet consortia in Compute Canada, 
and GridPP in the United Kingdom.
In addition, participation of individual researchers and institutions has been further 
supported by funds from ERC (FP7), H2020 Grant No. RISE-GA644294-JENNIFER, EU; 
JSPS, Japan; 
Royal Society, UK; 
and the DOE Early Career program, USA.


CNRS/IN2P3: Centre National de la Recherche Scientifique—Institut National
de Physique Nucle´aire et de Physique des Particules RSF: Russian Science
Foundation
MES: Ministry of Education and Science, Russia
ERDF: European Regional Development Fund
SNSF: Swiss National Science Foundation
SER (should be SERI): State Secretariat for Education, Research and
Innovation


\pagestyle{plain}
\clearpage


\bibliography{t2k2sens}
 
\end{document}